\documentclass{ieeeaccess}
\usepackage{cite}
\usepackage{amsmath,amssymb,amsfonts}
\usepackage{algorithmic}
\usepackage{graphicx}
\usepackage{textcomp}
\usepackage{multirow}
\usepackage{epstopdf}
\usepackage{bm}
\usepackage{psfrag}
\usepackage{pstool}
\usepackage[english]{babel}
\usepackage{graphicx}
\usepackage[absolute]{textpos}

\makeatletter
\AtBeginDocument{\DeclareMathVersion{bold}
\SetSymbolFont{operators}{bold}{T1}{times}{b}{n}
\SetSymbolFont{NewLetters}{bold}{T1}{times}{b}{it}
\SetMathAlphabet{\mathrm}{bold}{T1}{times}{b}{n}
\SetMathAlphabet{\mathit}{bold}{T1}{times}{b}{it}
\SetMathAlphabet{\mathbf}{bold}{T1}{times}{b}{n}
\SetMathAlphabet{\mathtt}{bold}{OT1}{pcr}{b}{n}
\SetSymbolFont{symbols}{bold}{OMS}{cmsy}{b}{n}
\renewcommand\boldmath{\@nomath\boldmath\mathversion{bold}}}
\makeatother

\def\BibTeX{{\rm B\kern-.05em{\sc i\kern-.025em b}\kern-.08em
    T\kern-.1667em\lower.7ex\hbox{E}\kern-.125emX}}

\newcommand{\copyrightstatement}{
    \begin{textblock}{13}(0.8,0.1)    
         \noindent
         \centering
         \textblockcolour{white}
         \footnotesize
         PREPRINT \copyright 2026 The Authors. This work is licensed under a Creative Commons Attribution-NonCommercial-NoDerivatives 4.0 License.
    \end{textblock}}

\begin{document}
\copyrightstatement
\history{Submitted 31 December 2025, accepted 14 February 2026, date of publication 9 March 2026, date of current version 12 March 2026.}
\doi{10.1109/ACCESS.2026.3672053}

\title{Energy-Aware Frame Rate Selection for Video Coding}
\author{\uppercase{Geetha Ramasubbu}\authorrefmark{1}, \IEEEmembership{Member, IEEE},
\uppercase{Andr\'e Kaup}\authorrefmark{1}, \IEEEmembership{Fellow, IEEE}, and Christian Herglotz\authorrefmark{2}, \IEEEmembership{Member, IEEE}}
\address[1]{Multimedia Communications and Signal Processing, Friedrich-Alexander University Erlangen-Nürnberg (FAU), 91058 Erlangen, Germany}
\address[2]{Chair of Computer Engineering, Brandenburgisch-Technische Universitat Cottbus-Senftenberg, 03046 Cottbus, Germany}
\tfootnote{This work was partially funded by the Deutsche Forschungsgemeinschaft (DFG, German Research Foundation), project number 447638564.}

\markboth
{Ramasubbu \headeretal: Energy-Aware Frame Rate Selection for Video Coding}
{Ramasubbu \headeretal: Energy-Aware Frame Rate Selection for Video Coding}

\corresp{Corresponding author: Geetha Ramasubbu (e-mail: geetha.ramasubbu@fau.de).}

\begin{abstract}
The demand for high-quality, immersive video experiences has necessitated adopting higher frame rates for more realistic scene portrayal. However, there is a growing demand for energy-saving video applications, where temporal downsampling is a crucial energy-saving factor. \textcolor{black}{To this end, the main contributions of this paper are twofold: First, we present an in-depth analysis of the impact of frame rate reductions on the visual quality of the video and the encoding as well as decoding energy. Second, we propose a lightweight frame rate selection method for energy- and quality-aware encoding. Concerning the first contribution, this paper performs extensive encoding and decoding measurements, followed by an investigation of the impact of temporal downsampling on the energy demand of encoding and decoding at different frame rates. Furthermore, we determine the objective visual quality of the downsampled videos. As a result of this investigation, we identify content- and quantization-setting-dependent energy-aware frame rates, i.e., the temporal downsampling factors that lead to Pareto-optimality in terms of energy and quality. We demonstrate that significant energy savings are achieved while maintaining constant visual quality. Subsequently, a subjective experiment is conducted to verify this observation regarding perceptual quality using mean opinion scores. As the second contribution, we propose an energy-aware frame rate selection method that extracts spatio-temporal features from the video sequences. Based on these features, the proposed method employs a feature-based supervised machine learning approach to predict energy-aware frame rates for a given quantization parameter and video sequence, aiming to reduce energy consumption during encoding and decoding.} The experimental results demonstrate that the proposed method offers significant energy savings, with an average of 17.46\% and 17.60\% of encoding and decoding energy demand reduction, respectively, alongside 3.38\% average bitrate savings at a constant quality.
\end{abstract}

\begin{keywords}
Video coding, energy efficiency, compression efficiency, frame rate selection, temporal downsampling
\end{keywords}

\titlepgskip=-21pt

\maketitle

\section{Introduction}
\label{sec:introduction}

\PARstart{T}{he} advent of portable devices, accessible and affordable Internet, mobile video streaming, and immersive and high-quality video experiences has significantly increased Internet video traffic in recent years. Moreover, immersive video demands expanding the dimensions of the videos beyond commonly used spatial and temporal resolutions, color gamut, and dynamic range. \textcolor{black}{As a consequence, there is an increasing volume of Internet and mobile video traffic \cite{globalInternetTraffic23}, exacerbating the increased bandwidth and energy demand. In addition, recent studies show that the data centers, which primarily use CPU-based software encoding \cite{Herglotz2022_SweetStreams}, currently account for approximately 3\% of global electricity consumption and is expected to increase to 8\% by 2030 \cite{HuaweiReport}. Although estimates differ, there is broad agreement on the severity of this emerging trend. Further, end-user devices account for a significant share of energy consumption in capturing, transmitting, and displaying videos \cite{Herglotz2022_SweetStreams}. In addition to the global impact, this poses a problem for battery-powered devices, as their batteries drain quickly due to increased energy requirements. Together, these factors emphasize the unsustainable environmental impact of online video and highlight the importance of integrating energy-aware strategies into the video encoding and streaming pipeline.}

Economically and environmentally-aware video streaming can be achieved with the development of powerful video compression techniques that aim for bitrate and energy efficiency without impairing the quality of experience (QoE). \textcolor{black}{In this paper, we exploit the observation that reducing the frame rate leads to substantial power savings \cite{Li12}, thereby increasing the operating time of battery-driven devices and reducing overall energy consumption. When video sequences are coded at the original frame rate, traditional energy-saving compression methods can only achieve a trade-off between quality and energy, as both quality degradation and energy consumption cannot be minimized simultaneously \cite{QOMEX2024}. However, when temporal variability is low in a video, human viewers might not need the original frame rate \cite{Song01}. Thus, adapting the frame rate in a sequence-specific manner enables energy-aware frame rate selection, i.e., selecting a frame rate that saves energy without compromising visual quality.}  

In this respect, we face the challenge of determining the energy-aware frame rate for video coding so that the encoding and decoding processes consume less energy with minimal or no degradation in quality. This paper examines energy demand at varying frame rates to develop a frame rate selection method for temporal downsampling to perform energy-aware video coding. \textcolor{black}{Although much work has been done in improving the rate-distortion efficiency for a fixed and variable frame rate \cite{ouEtal11}, \cite{Bulletal18}, and \cite{Lee2021}, studies examining the impact of temporal downsampling on energy consumption remain limited \cite{Herglotz23} and \cite{Ghasempour2024EnergyAwareSA} with most efforts restricted to decoding energy.} To this end, the significant contributions of our work are as follows:

\begin{itemize}

   \item \textcolor{black}{We investigate the combined impact of temporal downsampling and compression on energy consumption of encoders and decoders, demonstrating its content-dependent nature.}

   \item \textcolor{black}{We employ a subjective assessment to show that encoding a video at a lower frame rate can lead to significant energy savings while keeping the perceived quality.}

   \item \textcolor{black}{At last, we propose a novel energy-aware frame rate selection (EAFRS) method that extracts spatio-temporal features from video sequences and employs a supervised machine learning approach to predict the energy-aware frame rate for a given quantization parameter, minimizing energy consumption of encoders and decoders.}

\end{itemize}

The rest of the paper is organized as follows: First, Section \ref{sec:literature} briefly reviews available frame rate selection methods. Then, Section \ref{sec:analysis} demonstrates the effectiveness of frame rate downsampling by introducing our experimental setup, analyzing the impact of temporal downsampling and compression on energy consumption, and performing both objective and subjective quality assessments. Then, Section \ref{sec:method} presents the spatio-temporal features and the proposed frame rate selection method, followed by an evaluation of the proposed method in Section \ref{sec:evaluation}. Lastly, conclusions and future work directions are presented in Section \ref{sec:conclusion}.

\section{Literature Review}
\label{sec:literature}

In the literature, one can find that a large body of research deals with general considerations on the impact of spatial and temporal downsampling on the visual quality of a video signal. For example, a theoretical introduction on the spatial and temporal sampling of visual signals is given in \cite{Watson13}, which shows that depending on the content of a video, a lot of the information is irrelevant to human viewers. In particular, it is shown that high spatial and temporal frequencies are often not visible and, thus, not relevant to the end-user's perceived quality.  

Consequently, one can find much research on spatial and temporal downsampling and its impact on visual quality. For example, concerning spatial downsampling, Wang et al. \cite{Wang14} found that in rate-constrained environments, significant quality gains can be achieved when reducing the spatial resolution optimally. Similarly, Afonso et al. \cite{Afonso17} proposed a selection algorithm depending on the content, which achieved rate savings of up to $4\%$. Finally, \cite{Herglotz19,Dragic14}, shows that spatial downsampling leads to a significant amount of power savings on end-user devices. 

Regarding temporal downsampling, several attempts were made to implement a frame rate selection mechanism that gives an optimal frame rate for a sequence without affecting the quality. Ou et al. investigated the impact of frame rate and quantization on perceptual quality \cite{ouEtal11} and proposed the Q-STAR quality model as a function of spatial resolution, temporal resolution, and quantization step size in \cite{ouEtal12}. However, only frame rates up to 50 fps are considered \cite{ouEtal11}. Ma et al. proposed a bitrate model and a perceptual quality model for compressed videos as functions of frame rate and quantization step size \cite{Ma12}. A feature-based model that provides the optimal combination of frame rate and quantization step size for a given bitrate was introduced. However, they considered frame rates only up to 30 fps. In \cite{huangEtal16}, Huang et al. introduced a feature-based machine learning approach for frame rate selection. However, only sequences with frame rates up to 60 fps were used. Finally, in \cite{Bulletal18}, Katsenou et al. introduced a feature-based frame rate selection method. Even though sequences up to 120 fps were used, frame rate selection is limited to two frame rates, i.e.,  120 fps and 60 fps.  

In \cite{Mackin19}, Mackin et al. discuss the influences of higher frame rates and frame rate changes on the visual quality of the video sequences. The Mean Opinion Scores (MOS), a measure of visual quality obtained in \cite{Mackin19}, shows that increased frame rates lead to increased perceived visual quality. In addition, Mackin et al. find that the impact of frame rate changes on the visual quality is highly content-dependent. For example, a high impact of the frame rate on the visual quality is observed when the sequence inhibits large motion. Similarly, \cite{Herglotz20} studies the impact of frame rate reduction on objective quality and bitrate. The results show that depending on the sequence's content, reducing the frame rate is better than increasing the quantization step size. Furthermore, Herrou et al. propose a variable frame rate solution for significant bitrate and complexity reduction while preserving the visual quality of high frame rate (HFR) content \cite{variableFR21}.

Concerning the power consumption of end-user devices, previous studies have shown that it is beneficial to reduce the frame rate of a video sequence for power-saving applications \cite{Li12}, \cite{Yu15}, especially in the case of mobile devices. In addition, \cite{Herglotz19} shows that sequence-specific frame rates help to save on power. Even though scalable video coding offers temporal \cite{SHVC}, spatial, and quality scalabilities, it considers only compression efficiency. In addition, none of the prior works have attempted to investigate the collective impact of temporal downsampling and compression on energy consumption. Therefore, we study the collective impact of temporal downsampling and compression on energy demand and propose a frame rate selection method for energy-efficient video coding.

\textcolor{black}{Some of the works mentioned above perform frame rate selection and recommend a frame rate considering only the objective quality. In addition, very few works study the impact of temporal downsampling on decoding energy \cite{Herglotz23} and \cite{ Ghasempour2024EnergyAwareSA} but do not propose an energy-aware frame rate recommendation considering encoding and decoding energy. Therefore, our work proposes the EAFRS method that employs a feature-based supervised machine learning approach to perform energy-aware frame rate selection to aim for energy efficiency in encoder and decoder without sacrificing the video quality.}

\section{Temporal Downsampling Analysis}
\label{sec:analysis}

\textcolor{black}{This section begins with a brief introduction of the experimental setup in Section \ref{subsec:expSetup}, followed by an analysis of the joint impact of temporal downsampling and compression on energy consumption in Section \ref{subsec:energyEfficiency}. Then, Section \ref{subsec:optimalFr} explains how energy-aware frame rates for various CRF values are obtained from the energy-distortion curves, and in the end, Section \ref{subsec:subTest} presents a subjective evaluation of the energy-aware frame rate and CRF pairs from Section \ref{subsec:optimalFr}.}

\subsection{Experimental Setup}
\label{subsec:expSetup}
\begin{figure}
\centering
  \includegraphics[width=0.80\linewidth]{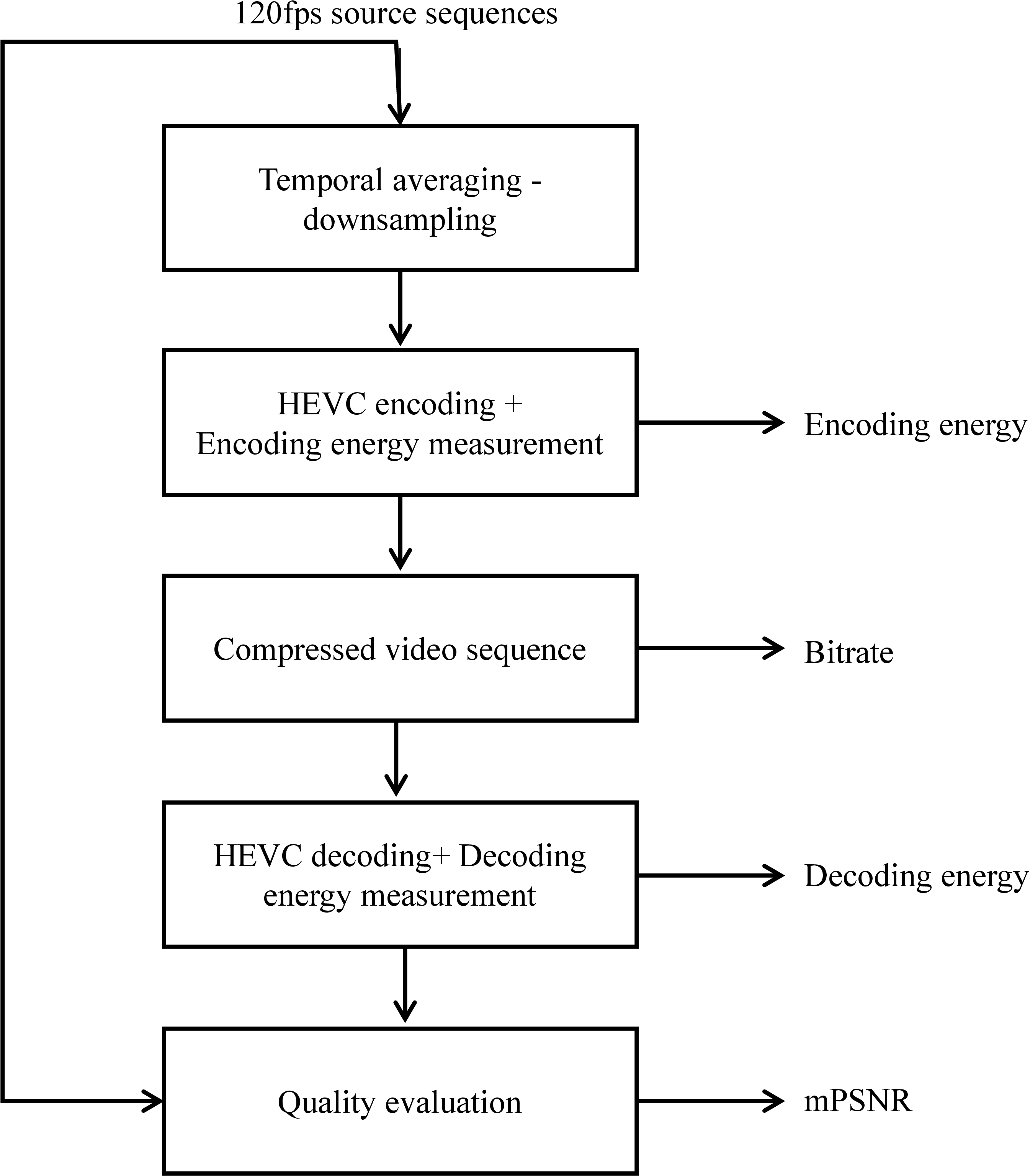}
  \caption{Experimental setup to study the impact of temporal downsampling on compression and energy efficiency.}
  \label{fig:setup}
\end{figure}

\textcolor{black}{An experimental setup, illustrated in Fig. \ref{fig:setup}, is designed to investigate the impact of temporal downsampling and compression on energy efficiency. For this purpose, we use 22 and six sequences with a high frame rate, i.e., 120 fps, from two publicly available datasets {BVI\_HFR} \cite{Mackin15}, and UVG \cite{Mercat2020}, respectively, as source sequences. Initially, these source sequences are temporally downsampled to lower frame rates (100, 60, 50, 40, 30, 25, 24, and 15 fps) using temporal averaging (frame averaging), which reduces temporal aliasing artifacts compared to frame dropping. The weights, for non-integer downsampling factors, are calculated by the weight generation algorithm introduced in \cite{Herglotz20}. For integer downsampling, weights are equal for all frames.}

\textcolor{black}{In the second step, these source sequences and their downsampled versions are encoded using an HEVC encoder wrapper, libx265 of FFmpeg \cite{FFmpeg}, with different Constant Rate Factor (CRF) values ranging from 0 to 51, in increments of 3. The bitrate of the encoded bit stream is recorded at the end of encoding. Then, we decode the encoded bit stream using the video decoder of FFmpeg (OpenHEVC).}

\textcolor{black}{In addition to encoding and decoding, we perform the corresponding energy measurements, as in general the encoding energy consumption correlates with the power and encoding time \cite{Ramasubbu22}, as a higher computational complexity leads to prolonged device activity and consequently higher energy use. However, other factors, such as CPU load distribution and dynamic voltage scaling, also influence energy independently of time alone. Therefore, direct measurements of energy, rather than only time, provide a more accurate evaluation. We describe the energy consumption of the encoding and decoding processes using two consecutive measurements each, as explained in \cite{Ramasubbu22} and \cite{Herglotz18}. For the encoding process, the first measures the total energy consumption of the device during the encoding process as follows \cite{Ramasubbu22}:}

\begin{equation} \label{eq:totalenergy}E_{\mathrm {enc, total}} = \int _{t=0}^{T_{\mathrm{enc}}} P_{\mathrm {enc, total}}(t) \mathrm{d}t, \end{equation}
where $T_{\mathrm{enc}}$ is the duration of the encoding process and $P_{\mathrm {enc, total}}(t)$ is the total power consumption of the device while encoding. \textcolor{black}{This is followed by the second measurement, which measures the energy consumed over the same encoding time duration $T_{\mathrm{enc}}$, in idle mode, and is given as follows \cite{Herglotz18}:}
\begin{equation} \label{eq:idle}E_{\mathrm {enc,   idle}} = \int _{t=0}^{T_{\mathrm{enc}}} P_{\mathrm {enc, idle}}(t) \mathrm{d}t, \end{equation}
where $P_{\mathrm {enc, idle}}(t)$ is the power consumed by the device in idle mode.
\textcolor{black}{Ultimately, we describe the encoding energy as the difference between the two measurements from Eqs. \eqref{eq:totalenergy} and \eqref{eq:idle} as follows \cite{Herglotz18}:}
\begin{equation} E_{\mathrm {enc}} = E_{\mathrm {enc, total}}- E_{\mathrm {enc, idle}}. \end{equation}
Similarly, we measure the decoding energy using two consecutive measurements \cite{Herglotz18} and is given as follows: 
\begin{equation} E_{\mathrm {dec}} = \int _{t=0}^{T_{\mathrm{dec}}} P_{\mathrm {dec, total}}(t) \mathrm{d}t- \int _{t=0}^{T_{\mathrm{dec}}} P_{\mathrm {dec, idle}}(t) \mathrm{d}t. \end{equation}

\textcolor{black}{In this work, we performed energy measurements on an  Intel i7-7500 CPU with eight cores, where we employed the integrated power meter in Intel CPUs, running average power limit (RAPL) \cite{RAPLinAction2018}, that directly returns aggregated energy values such as $E_{\mathrm {enc, total}}$ and $E_{\mathrm {enc, idle}}$. In addition, we repeat each measurement multiple times and perform a confidence interval test proposed in \cite{Bendat1971} to assess the statistical significance of the measured encoding energies \cite{Ramasubbu22} and decoding energies \cite{Herglotz17a}, thereby avoiding the influence of noise and background processes. It should be noted that we use the encoding and decoding process alone for energy measurements and do not include spatio-temporal feature extraction.}

After decoding, the quality evaluation is performed using the 120 fps source sequence as the reference. \textcolor{black}{To account for non-integer downsampling factors in this study, the matched quality evaluation, matched Peak Signal-to-Noise Ratio (mPSNR), an extension of traditional Peak Signal-to-Noise Ratio (PSNR), introduced in \cite{Herglotz20} is used. It introduces virtual frames at the least common multiple (LCM) of the original and downsampled frame rates, ensuring a fair comparison between frames that lack a one-to-one mapping in the time domain. As noted in \cite{Herglotz20}, PSNR and mPSNR yield the same value for any integer downsampling factors.}

\textcolor{black}{Lastly, we use Bjøntegaard Delta (BD) \cite{Bjoentegaard01} metrics to evaluate differences in rate-distortion and energy-distortion performance across different frame rates. The Bjøntegaard Delta Rate (BDR) quantifies the average bitrate difference at a constant visual quality. In contrast, the Bjøntegaard Delta Encoding Energy (BDEE) and Bjøntegaard Delta Decoding Energy (BDDE) are average energy differences at the same quality for the encoding and decoding processes, respectively. In addition, all BD values are computed following the improved interpolation method proposed in \cite{HerglotzBB2024}, which enhances the precision of BDR and energy-related BD metrics. Negative BD values indicate a reduction in bitrate, encoding energy, or decoding energy, thereby reflecting improved efficiency.}

\subsection{Temporal Downsampling and Energy Demand}
\label{subsec:energyEfficiency}
Energy-distortion curves visualize the energy-quality trade-off. Therefore, the energy-distortion curve is crucial for selecting the optimal encoding parameters and picking a trade-off between energy and distortion. By including an additional parameter, i.e., frame rate, in the energy-distortion curve, we can study the impact of temporal downsampling on both energy and quality. As an example, the encoding energy-distortion curves of all the considered frame rates $f\in \{120, 100, 60, 50, 40, 30, 25, 24, 15\}$ for a sequence of the {BVI\_HFR} video dataset are illustrated in Figure \ref{fig:catch_pareto}. 

\textcolor{black}{Each colored line in the energy-distortion curves of Figure \ref{fig:catch_pareto} represents the energy-distortion curve for a single frame rate. Each marker indicates the quality and energy for CRF values ranging from 0 to 51 (with a step size of 3), from the top right (higher quality and higher energy demand) to the bottom left (lower quality and lower energy demand). A specific frame rate-CRF pair is Pareto-optimal if no tested pair achieves a higher mPSNR with the same or lower energy, or lower energy with the same or higher mPSNR. As a result, the Pareto-optimal points represent the best attainable trade-offs. For example, at high visual quality (above 42 dB), the native frame rate (120 fps) should be used. For low qualities (below 36 dB), the lowest tested frame rate of 15  fps (cyan curve) is optimal. Collectively, all sets of Pareto-optimal points form the Pareto front in the energy distortion plane. The Pareto front thus supports selecting optimal encoding parameters by identifying the Pareto-optimal frame rate-CRF pairs across the explored operating range.}

The observations from all the energy-distortion curves can be summarized as follows: For certain sequences, such as the 'catch' (Fig. \ref{fig:catch_pareto}), 'flowers,' 'gold\_side,' 'lamppost,' and 'pond' sequences, the first intersection of the energy-distortion curves occurs at low CRF values (intersection of the curves for 120 fps and 60 fps as in Fig. \ref{fig:catch_pareto}). \textcolor{black}{Consequently, significant energy reductions can be achieved at high objective qualities by temporal downsampling. Further content analysis reveals that these sequences commonly exhibit a static background and motion confined to specific regions of the video.}

For sequences with motion across all frames, such as 'bobblehead,' 'hamster,' and 'water\_splashing,' the first intersection occurs at higher CRF values, leading to strong quality degradation from temporal downsampling, \textcolor{black}{limiting the achievable energy reduction at high qualities. Hence, energy reductions are possible only at lower objective qualities by temporal downsampling.} For the sequences with dynamic structures, i.e., spatially irregular structures moving as a continuum, such as 'water\_ripples', the first intersection can be observed at intermediate CRF values. \textcolor{black}{As a result, energy reductions can only be achieved at intermediate objective qualities.}

The decoding energy-distortion curves exhibit behavior similar to that of the encoding energy-distortion curves. Summarizing, we find that the impact of temporal downsampling on encoding and decoding energy is content-dependent. Therefore, we can exploit this behavior to enable energy-aware video encoding and decoding without compromising video quality in a content-dependent manner. 

\subsection{Energy-Aware Temporal Downsampling}
\label{subsec:optimalFr}
\begin{figure}[t]
  \includegraphics[width=0.9\linewidth]{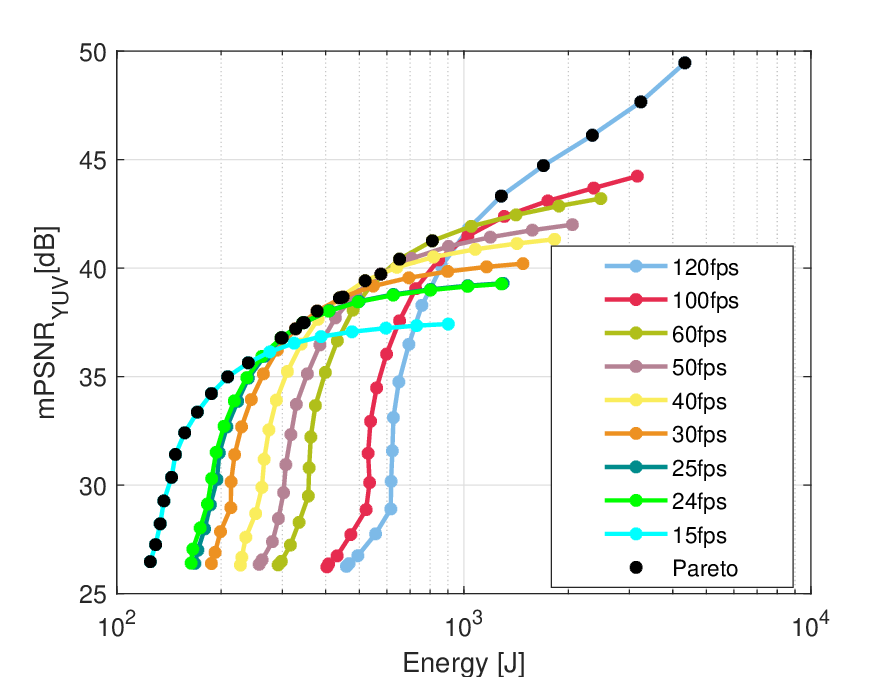}
  \caption{Impact of temporal downsampling on encoding energy and objective quality for the `catch' sequence of {BVI\_HFR}.}
  \label{fig:catch_pareto}
\end{figure}

\textcolor{black}{To facilitate energy reduction, the energy-distortion curves (discussed in Subsection \ref{subsec:energyEfficiency}) are used to identify the energy-aware frame rate for a particular CRF. For the sake of simplicity, we consider the commonly used CRF values: $c \in \{18, 23, 28, 33\}$ \cite{FFmpeg}. The first step is to obtain the Pareto-optimal points that are energy-efficient, which are obtained from the energy-distortion curves as illustrated in Figure \ref{fig:catch_pareto}. Then, four energy-aware frame rates for the given subset of CRF values are obtained as follows:}
\begin{enumerate}
     \item \textcolor{black}{For the lowest CRF value of 18, we take the frame rate corresponding to the highest objective quality as the energy-aware frame rate $f_{18}$, which is, in most cases, the source sequences' frame rate.}
    \item \textcolor{black}{For the next CRF value, first, we take all Pareto-efficient points. From this set, we pick the energy-aware frame rate as the frame rate with the smallest distortion difference to the point corresponding to the previous energy-aware frame rate and the current CRF.}
    \item Repeat step 2 for the remaining CRF values.
\end{enumerate}
\textcolor{black}{Table \ref{tab:optFR} summarizes these energy-aware frame rates for the corresponding crf values, bitrate savings, and energy savings achieved through energy-aware temporal downsampling for all video sequences considered in this work. To calculate BD values, the reference is compression with all CRF values at the original frame rate. The sequences with motion across all frames, such as bobblehead and books, show no energy savings, as maintaining the maximum frame rate is necessary to preserve quality. In contrast, sequences with static backgrounds and localized motion, such as pond and honeybee, achieve significant energy reductions, with encoding and decoding energy savings reaching approximately 85\% and 83\%, respectively. These sequences benefit from temporal downsampling even at high objective qualities. For sequences with dynamic structures characterized by spatially irregular patterns moving as a continuum, such as water\_ripples, moderate energy savings are observed. On average, considering all sequences, the observed savings are 2.68\% in bitrate, 19.45\% in encoding energy, and 19.53\% in decoding energy. When focusing only on sequences where frame rate reduction is beneficial, the average savings increase to 4.41\%, 32.03\%, and 32.17\%, respectively.}

\textcolor{black}{Moreover, Table \ref{tab:optFR} shows that the energy-aware frame rates consistently correspond to integer downsampled versions of the original frame rate. The energy-aware frame rates listed in column 2 will serve as ground-truth labels for training the proposed EAFRS method (see Section VI).}

\begin{table}[]
\centering
\caption{Energy-aware frame rates (ground truth), \{$f_{18}$,$f_{23}$,$f_{28}$,$f_{33}$\}, for the subset of CRF values $c \in \{18, 23, 28, 33\}$, with their associated BDR, BDEE, and BDDE values, which represents the average bitrate, encoding, and decoding energy savings for all the sequences.}
\begin{tabular}{|l|lrrr|}
\hline
 & \multicolumn{4}{c|}{\textbf{Energy-Aware Temporal Downsampling}} \\ \hline
\multirow{2}{*}{\textbf{File Name}} & \multicolumn{1}{l|}{\multirow{2}{*}{\begin{tabular}[c]{@{}l@{}}\textbf{Energy-aware}\\ \textbf{frame rates}\end{tabular}}} & \multicolumn{3}{c|}{\textbf{BD (\%)}} \\ \cline{3-5}  & \multicolumn{1}{l|}{} & \multicolumn{1}{l|}{\textbf{BDR}} & \multicolumn{1}{l|}{\textbf{BDEE}} & \textbf{BDDE} \\ \hline
bobblehead & \multicolumn{1}{l|}{\{120,120,120,120\}} & \multicolumn{1}{r|}{0} & \multicolumn{1}{r|}{0} & 0 \\ \hline
books & \multicolumn{1}{l|}{\{120,120,120,120\}} & \multicolumn{1}{r|}{0} & \multicolumn{1}{r|}{0} & 0 \\ \hline
bouncyball & \multicolumn{1}{l|}{\{120,120,120,120\}} & \multicolumn{1}{r|}{0} & \multicolumn{1}{r|}{0} & 0  \\ \hline
catch & \multicolumn{1}{l|}{\{120,30,15,15\}} & \multicolumn{1}{r|}{-16.03} & \multicolumn{1}{r|}{-69.45} & -64.60  \\ \hline
catch\_track & \multicolumn{1}{l|}{\{120,120,120,120\}} & \multicolumn{1}{r|}{0} & \multicolumn{1}{r|}{0} & 0 \\ \hline
cyclist & \multicolumn{1}{l|}{\{120,120,120,120\}} & \multicolumn{1}{r|}{0} & \multicolumn{1}{r|}{0} & 0 \\ \hline
flowers & \multicolumn{1}{l|}{\{120,30,15,15\}} & \multicolumn{1}{r|}{16.43} & \multicolumn{1}{r|}{-52.73} & -52.61 \\ \hline
golf\_side & \multicolumn{1}{l|}{\{120,15,15,15\}} & \multicolumn{1}{r|}{-14.75} & \multicolumn{1}{r|}{-77.37} & -73.10 \\ \hline
guitar\_focus & \multicolumn{1}{l|}{\{120,120,30,24\}} & \multicolumn{1}{r|}{13.73} & \multicolumn{1}{r|}{-23.72} & -27.12 \\ \hline
hamster & \multicolumn{1}{l|}{\{120,120,120,120\}} & \multicolumn{1}{r|}{0} & \multicolumn{1}{r|}{0} & 0 \\ \hline
joggers & \multicolumn{1}{l|}{\{120,120,120,15\}} & \multicolumn{1}{r|}{1.84} & \multicolumn{1}{r|}{-5.07} & -5.91 \\ \hline
lamppost & \multicolumn{1}{l|}{\{120,60,15,15\}} & \multicolumn{1}{r|}{20.04} & \multicolumn{1}{r|}{-32.08} & -32.58 \\ \hline
leaves\_wall & \multicolumn{1}{l|}{\{120,120,24,15\}} & \multicolumn{1}{r|}{0.33} & \multicolumn{1}{r|}{-20.14} & -20.97 \\ \hline
library & \multicolumn{1}{l|}{\{120,120,120,15\}} & \multicolumn{1}{r|}{-0.68} & \multicolumn{1}{r|}{-4.46} & -4.48 \\ \hline
martial\_arts & \multicolumn{1}{l|}{\{120,120,120,30\}} & \multicolumn{1}{r|}{0.82} & \multicolumn{1}{r|}{-3.48} & -4.37 \\ \hline
plasma & \multicolumn{1}{l|}{\{120,120,120,120\}} & \multicolumn{1}{r|}{0} & \multicolumn{1}{r|}{0} & 0 \\ \hline
pond & \multicolumn{1}{l|}{\{60,15,15,15\}} & \multicolumn{1}{r|}{-55.85} & \multicolumn{1}{r|}{-82.54} & -79.96 \\ \hline
pour & \multicolumn{1}{l|}{\{120,120,120,30\}} & \multicolumn{1}{r|}{0.21} & \multicolumn{1}{r|}{-3.57} & -4.86 \\ \hline
sparkler & \multicolumn{1}{l|}{\{120,120,120,120\}} & \multicolumn{1}{r|}{0} & \multicolumn{1}{r|}{0} & 0 \\ \hline
typing & \multicolumn{1}{l|}{\{120,120,24,15\}} & \multicolumn{1}{r|}{1.09} & \multicolumn{1}{r|}{-30.27} & -33.09 \\ \hline
water\_ripples & \multicolumn{1}{l|}{\{120,120,24,24\}} & \multicolumn{1}{r|}{16.27} & \multicolumn{1}{r|}{-17.21} & -19.12 \\ \hline
water\_splashing & \multicolumn{1}{l|}{\{120,120,120,120\}} & \multicolumn{1}{r|}{0} & \multicolumn{1}{r|}{0} & 0 \\ \hline
Beauty & \multicolumn{1}{l|}{\{120,120,120,30\}} & \multicolumn{1}{r|}{1.52} & \multicolumn{1}{r|}{-4.12} & -5.50 \\ \hline
Bosphorus & \multicolumn{1}{l|}{\{120,120,30,30\}} & \multicolumn{1}{r|}{9.43} & \multicolumn{1}{r|}{-30.18} & -32.09 \\ \hline
HoneyBee & \multicolumn{1}{l|}{\{30,15,15,15\}} & \multicolumn{1}{r|}{-71.71} & \multicolumn{1}{r|}{-85.10} & -82.78 \\ \hline
Jockey & \multicolumn{1}{l|}{\{120,120,120,120\}} & \multicolumn{1}{r|}{0} & \multicolumn{1}{r|}{0} & 0 \\ \hline
ReadySteadyGo & \multicolumn{1}{l|}{\{120,120,120,120\}} & \multicolumn{1}{r|}{0} & \multicolumn{1}{r|}{0} & 0 \\ \hline
YachtRide & \multicolumn{1}{l|}{\{120,120,120,30\}} & \multicolumn{1}{r|}{2.28} & \multicolumn{1}{r|}{-3.00} & -3.76 \\ \hline
\textbf{All sequences} & \multicolumn{1}{l|}{\textbf{Average BD}} & \multicolumn{1}{r|}{\textbf{-2.68}} & \multicolumn{1}{r|}{\textbf{-19.45}} & \textbf{-19.53} \\ \hline
\textbf{Downsampled} & \multicolumn{1}{l|}{\textbf{Average BD}} & \multicolumn{1}{r|}{\textbf{-4.41}} & \multicolumn{1}{r|}{\textbf{-32.03}} & \textbf{-32.17} \\ 
\textbf{sequences} & \multicolumn{1}{l|}{} & \multicolumn{1}{r|}{} & \multicolumn{1}{r|}{} &  \\ \hline
\end{tabular}
\label{tab:optFR}
\end{table}

\subsection{Subjective Evaluation}
\label{subsec:subTest}
To show that downsampling is also beneficial in terms of perceptual quality, we performed a subjective assessment of the sequences encoded with the (ground truth) energy-aware frame rate and CRF pairs (obtained in Section \ref{subsec:optimalFr}) and the sequences encoded with the fixed source frame rate and CRF pairs. \textcolor{black}{With this assessment, we can determine the correlation between the objective quality (mPSNR) and subjective evaluation. Consequently, we validate that temporal downsampling, when applied selectively based on content, helps preserve visual quality while reducing energy consumption.}

We conducted the subjective test using a calibrated Fujitsu LCD monitor with ($1920 \times 1080$ resolution, $60\,$Hz, 300 $\mathrm{cd/m^2}$ peak luminance, and 1000:1 as static contrast ratio) and a 150 cm viewing distance, conforming to the conditions mentioned in BT.500-14 \cite{ITU_BT500_2019}. This setup was powered by a PC with Matlab for test control and FFmpeg \cite{FFmpeg} for playback. Prior to the session, participants were screened for corrected-to-normal or normal visual acuity using the Snellen chart and for normal color vision using the Ishihara charts \cite{ITU_P910_2008}. All 16 participants (eight experts and eight non-experts) met the criteria, with no errors on 20/30 line and no more than two errors on 12 Ishihara plates.

In this work, the Absolute Category Rating (ACR) \cite{ITU_P910_2008} was used, in which the test sequences were presented individually and rated independently on a five-level category scale. After each sequence, participants evaluated its quality within $10 \mathrm{s}$, followed by a $5 \mathrm{s}$ mid-level grey screen before the next sequence. The total duration of the subjective test for each participant was limited to 30 minutes. The subjective evaluation is performed on sequences where downsampling below $60\,$Hz saves bitrate and energy, comprising the 'catch,' 'flowers,' 'golf\_side,' 'pond,' and 'typing' sequences. For these video sequences, both those encoded with optimal frame rate/CRF pairs and those encoded with fixed source frame rate/CRF pairs were presented for subjective testing. In addition, the sequences were displayed to each participant in a random and unique order. 

For each sequence, the Mean Opinion Score (MOS) is obtained by averaging the ratings of all participants. Then, the MOS obtained in the subjective assessment is evaluated against the objective quality mPSNR using the Spearman rank-order correlation coefficient (SROCC), which measures the strength and direction of association between two ranked variables \cite{Cowan1998}. The SROCC values for the sequences under subjective testing are tabulated in Table \ref{tab:srocccomparison}.
\begin{table}[]
    \centering
    \caption{The statistic correlation between the subjective MOS scores and objective quality {mPSNR}.}
    \label{tab:srocccomparison}
    \begin{tabular}{|c|c|}
    \hline
     Sequence Name    &  SROCC\\
     \hline \hline
     catch     & 0.89\\
     \hline
     flowers   & 0.82\\
     \hline
     golf\_side & 0.91\\
     \hline
     pond   & 0.60\\
     \hline
     typing  & 0.91\\
     \hline
    \end{tabular}
\end{table}

As Spearman correlation is a nonparametric measure, the exact relation between the compared variables is obtained without knowledge of their joint probability distribution. When the variables being compared are perfectly monotonically related, the Spearman correlation coefficient equals 1, and the sign depends on whether the relation is increasing or decreasing. In our case, the SROCC shows that the relationship between objective mPSNR and subjective MOS scores is monotonically increasing.

For the sequences 'golf\_side,' 'typing,' and 'catch,' the Spearman coefficient is 0.91, 0.91, and 0.89, respectively, suggesting a monotonically increasing relation between their mPSNR and MOS. \textcolor{black}{For the 'flowers' sequence, the Spearman coefficient is 0.82, which is less than that of the 'golf\_side,' 'typing,' and 'catch' sequences, which can be attributed to its content characteristics.}

\textcolor{black}{For the 'pond' sequence, the SROCC drops further to 0.60, indicating a weaker correlation between mPSNR and MOS. Again, the content characteristics explain this behavior. While the background is largely static, it contains multiple leaf-like structures with subtle motion. In such cases, where the background is almost completely static and fine details are present, viewers may focus more intently on fine-structure variations, while objective metrics such as mPSNR, which operate frame-wise and do not prioritize spatial attention, fail to capture these perceptual subtleties.} 

In general, we find that the high correlation indicates that the mPSNR is a useful first approximation of subjective visual quality. Therefore, we will use this metric to optimize downsampling decisions without compromising the quality and use it for the proposed frame rate selection method. 

\section{Proposed Energy-aware Frame Rate Selection Method}
\label{sec:method}
\begin{figure}[t]
\centering
  \includegraphics[width=7cm]{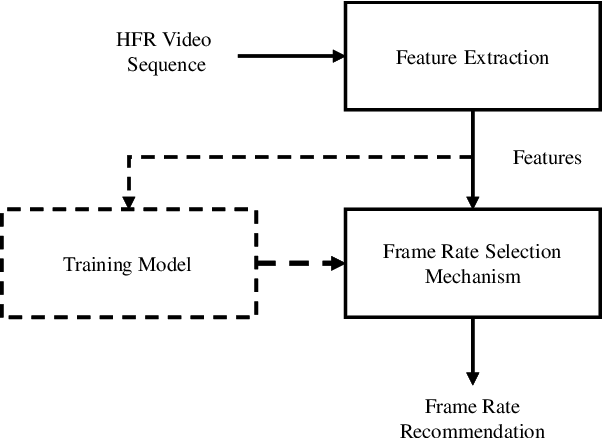}
  \caption{A diagrammatic illustration of the proposed EAFRS method.}
  \label{fig:selectionMechanism}
\end{figure}

This section describes the proposed energy-aware frame rate selection (EAFRS) method, which recommends a frame rate for a given CRF value to reduce energy demand for encoding and decoding. An illustration of the proposed method can be seen in Figure \ref{fig:selectionMechanism}. The content of the sequence is analyzed for features that represent its spatial and temporal properties to determine the energy-aware frame rate. The features used for this paper are discussed in Section \ref{subsec:featureExtraction}. Afterward, these features are exploited to select the energy-aware frame rate. The selection mechanism is explained in Section \ref{subsec:selectionMethod}. The proposed frame rate selection mechanism comprises two modes: training and test. We use the energy-aware downsampling frame rates from Section \ref{subsec:optimalFr} as ground truth, extracted spatiotemporal features, and CRF value to train the model in the training mode. Once the model is trained, we can use the selection mechanism in test mode to validate its accuracy and make energy-aware frame rate recommendations.

\subsection{Feature Extraction}
\label{subsec:featureExtraction}
As mentioned in  Section \ref{sec:analysis}, the impact of temporal downsampling and compression on encoding and decoding energy is content-dependent. As a result, features that categorize the video sequences based on their content are needed to enable content-dependent energy-aware frame rate selection. Here, the content of the video sequence includes characteristics such as motion complexity, motion type (e.g., translational, rotational, affine), motion speed, and spatial contrast (e.g., motion homogeneity, motion distribution) \cite{huangEtal16}. Such characterization has relevance to the human visual system as follows:

\begin{itemize}
    \item Reducing the frame rate causes a judder effect in fast-motion sequences and depends on speed \cite{huangEtal16}.
    \item Human eyes are more attentive towards moving objects. In general, we can distinguish object motion and camera motion, and both can be present in video sequences, described by the type of motion  \cite{huangEtal16}, \cite{Mackin17}.
    \item In addition, the human visual system has reduced capability to perceive edges, movements, and distortions in complex spatial backgrounds, which is distinguished by spatial contrast \cite{Mackin17}.
\end{itemize}

The first feature is the Frame difference (FD), a measure of temporal variation in a video, obtained from the absolute difference between co-located pixels in successive frames \cite{ouEtal11}. In addition, we introduce Squared frame difference (SFD), a temporal feature that is the ratio of the sum of the squared absolute difference between co-located pixels in successive frames to the product of the total number of frames $N$, width $W$, and height $H$,
\begin{equation}
\label{SFD}
    \mathrm{SFD} = \frac{\sum_{n=1}^{N}\sum_{w=1}^{W} \sum_{h=1}^{H}{|(F_{w,h,i+1}-F_{w,h,i})|}^2}{N\cdot W \cdot H},
\end{equation}
where $F_{w,h,i}$ is the pixel value for the given frame $i$ at given horizontal and vertical location $w,h$. Another temporal feature is optical flow (OF), which provides the distribution of apparent object velocities in the video. OF is computed based on Farneback's method \cite{farnbaeck03}. Furthermore, OF descriptors such as magnitude (mag) and orientation (or) of OF vectors are obtained. These descriptors characterize and distinguish the dynamic textures in the video. For example, spatially irregular structures \cite{bull17}, moving as a continuum (such as water, smoke), have a different OF pattern in comparison with spatially regular or irregular structures   \cite{bull17} with moving independent structures (such as leaves moving in the wind). 

Standard Deviation (STD) is a spatial feature that gives the average standard deviation of the pixel values in each frame and is used to measure the contrast of a video \cite{ouEtal11}. Another spatial feature is the Gray Level Co-occurrence Matrix (GLCM), which captures the intensity contrast between neighboring pixels in a frame \cite{bull17}. Therefore, GLCM captures the texture's directionality and degree of coarseness. From the GLCM, several descriptors, such as contrast, homogeneity, correlation, energy, and entropy, can be derived at the frame level \cite{bull17}. In addition, we used Spatial Information (SI) and Temporal Information (TI) as defined in \cite{ITU_P910_2008}, which are widely used features that approximate scene complexity. SI indicates the amount of spatial detail in an image, which is higher for more spatially complex scenes \cite{ITU_P910_2008}. TI quantifies the temporal changes in a video sequence, which are higher in high-motion sequences \cite{ITU_P910_2008}.

Furthermore, the HVS has reduced capability to perceive edges, motions, and distortions in complex spatial and temporal backgrounds \cite{ouEtal12}. Consequently, the combination of spatial and temporal attributes, such as the Histogram of Oriented Gradients (HoG), can be used to characterize the complexity of spatial and temporal backgrounds, known as spatio-temporal features. Normalized Frame Difference (NFD) is a spatiotemporal feature used to identify frame differences and to link contrast to motion between successive frames \cite{Bulletal18}. Lastly, we used low-processing-complexity features, average spatial energy (SE) calculated using a DCT-based energy function, and average temporal energy (TE) calculated based on block-wise SAD of the texture energy \cite{VCA}. Furthermore, we use the CRF  as a feature to differentiate between different quantization settings. A comprehensive list of the features used in this work, along with their type and their statistics' notation, is tabulated in Table \ref{tab:spatiotemporalfeatures}.
\begin{table}[]
    \centering
    \caption{List of spatio-temporal features considered in this work, with the features selected based on chi-square scores \cite{chiSquares} highlighted in bold.}
    \label{tab:spatiotemporalfeatures}
    \begin{tabular}{|c|c|c|}
    \hline
    \textbf{Feature} & \textbf{Feature Type} &  \textbf{Notation of feature Statistics}\\
    \hline \hline
    \textbf{FD} & Temporal & $\boldsymbol{\mathrm{meanFD}}$\\
    \hline
    \textbf{SFD} & Temporal & $\boldsymbol{\mathrm{meanSFD}}$\\
    \hline
    \textbf{STD}  & Spatial & $\mathrm{meanSTD}$\\
    \hline
    \textbf{SI}  & Spatial & $\boldsymbol{\mathrm{maxSI}}$\\
    \hline
    \textbf{TI}  & Temporal & $\boldsymbol{\mathrm{maxTI}}$\\
    \hline
    \textbf{GLCM} & Spatial & $\boldsymbol{\mathrm{meanGLCM_{con}}}$, $\mathrm{stdGLCM_{con}}$\\
    && $\mathrm{meanGLCM_{corr}}$, $\boldsymbol{\mathrm{stdGLCM_{corr}}}$ \\
    && $\mathrm{meanGLCM_{ene}}$, $\boldsymbol{\mathrm{stdGLCM_{ene}}}$\\
    && $\boldsymbol{\mathrm{meanGLCM_{hom}}}$, $\mathrm{stdGLCM_{hom}}$\\
    && $\boldsymbol{\mathrm{meanGLCM_{ent}}}$, $\boldsymbol{\mathrm{stdGLCM_{ent}}}$\\
    \hline
    \textbf{OF} & Temporal & $\boldsymbol{\mathrm{meanOF_{mag}}}$, $\mathrm{stdOF_{mag}}$\\
    &&$\mathrm{meanOF_{or}}$, $\mathrm{stdOF_{or}}$\\
    \hline
    \textbf{HoG} & Spatio-temporal & $\mathrm{meanHoG, StdHoG}$\\
    \hline
    \textbf{NFD} & Spatio-temporal & $\boldsymbol{\mathrm{meanNFD}}$\\
    \hline
    \textbf{SE} & Spatial & $\boldsymbol{\mathrm{meanE}}$\\
     \hline
    \textbf{TE} & Temporal & $\boldsymbol{\mathrm{meanh}}$\\
    \hline
    \textbf{CRF} & Encoding & $\boldsymbol{\mathrm{CRF}}$\\
    \hline
    \end{tabular}
\end{table}
\subsection{Frame Rate Selection as a classification problem}
\label{subsec:selectionMethod}
A supervised learning algorithm observes the training data (example input-output pairs) and produces an inferred function to map new samples. For example, given the training input-output pairs $(x_1, y_1),(x_2, y_2),...(x_n, y_n )$, also called features, each $y_j$, $j\in\{1,2,...,n\}$ is generated by an unknown function $y = f(x)$, and the task of supervised learning is to find a function $h(x)$ that approximates the true function $f(x)$ \cite{russell2016}. The learning problem is classification, where the output $y$ is one of a finite set of values, i.e., mapping to a distinct set of frame rates. 

Although the frame rate selection problem can be considered a regression problem, we have treated it as a classification problem in this work. The reason is twofold: First, standardization activities commonly use a discrete set of frame rates. Second, we tested the non-integer downsampling factors in the intermediate studies, and from Table \ref{tab:optFR}, we observe that the non-integer downsampled frame rates are not energy-efficient. Due to these reasons, it is more efficient to treat it as a classification problem.

The training dataset is used to find an optimal mapping for each target class. Once the mapping is determined, the next task is to predict the target class \cite{russell2016}. A classification task with two possible outcomes is called binary classification, whereas a classification task where each sample is mapped to one of many (more than two) classes is called multiclass classification \cite{libsvm}. The classes are mutually exclusive in both classification methods, so each sample can be labeled with only one class \cite{scikit-learn}. For frame rate selection, we use multiclass classification: the input sequences with specific spatiotemporal properties and desired input CRF values are classified into five classes, with the output corresponding to the optimal frame rate. The five output classes represent the frame rates $f \in \{120, 60, 30, 24, 15\}$. This work used ensemble learning \cite{hastie01statisticallearning}, a technique in which multiple base classifiers are generated, and a new classifier is derived from them that performs better than the constituent base classifiers. We use ensemble methods because the predictions of learning-based models can be adversely affected by bias, variance, and noise.

\section{Evaluation}
\label{sec:evaluation}
This section introduces feature selection, followed by the training and testing of the classification method in Section \ref{subsec:TT}. In the end, the results are discussed in terms of bitrate savings and energy savings. Furthermore, we compare the proposed method with a method from the literature in Section \ref{subsec:becnhmarking} and assess its complexity in Section \ref{subsec:complexity}.

\subsection{Feature Selection}
\label{subsec:selection}
For robust, low-complex classification, we reduce the dimensionality of the feature space by selecting a suitable subset of features. Feature selection is performed to avoid overfitting by modeling with an excessive number of features. In addition, feature selection reduces model size, improving computational performance and enabling deployment on memory-restricted devices. Lastly, feature selection improves model interpretability by using fewer features, which may help identify the features that most affect the model's behavior. 

Given a high correlation among features, different feature subsets can yield the same results. However, to achieve a low-complexity solution, we are interested in identifying a low-cardinality subset of features. Therefore, we employ a univariate feature ranking for classification using chi-square tests \cite{chiSquares}, which examine whether each predictor variable is independent of a response variable using individual chi-square tests. A larger chi-square score indicates that the corresponding predictor is significant \cite{chiSquares2}. The features are ranked based on chi-square scores, and this ranking is then used to select the optimal features.

\begin{figure*}[]
\includegraphics[width=\linewidth]{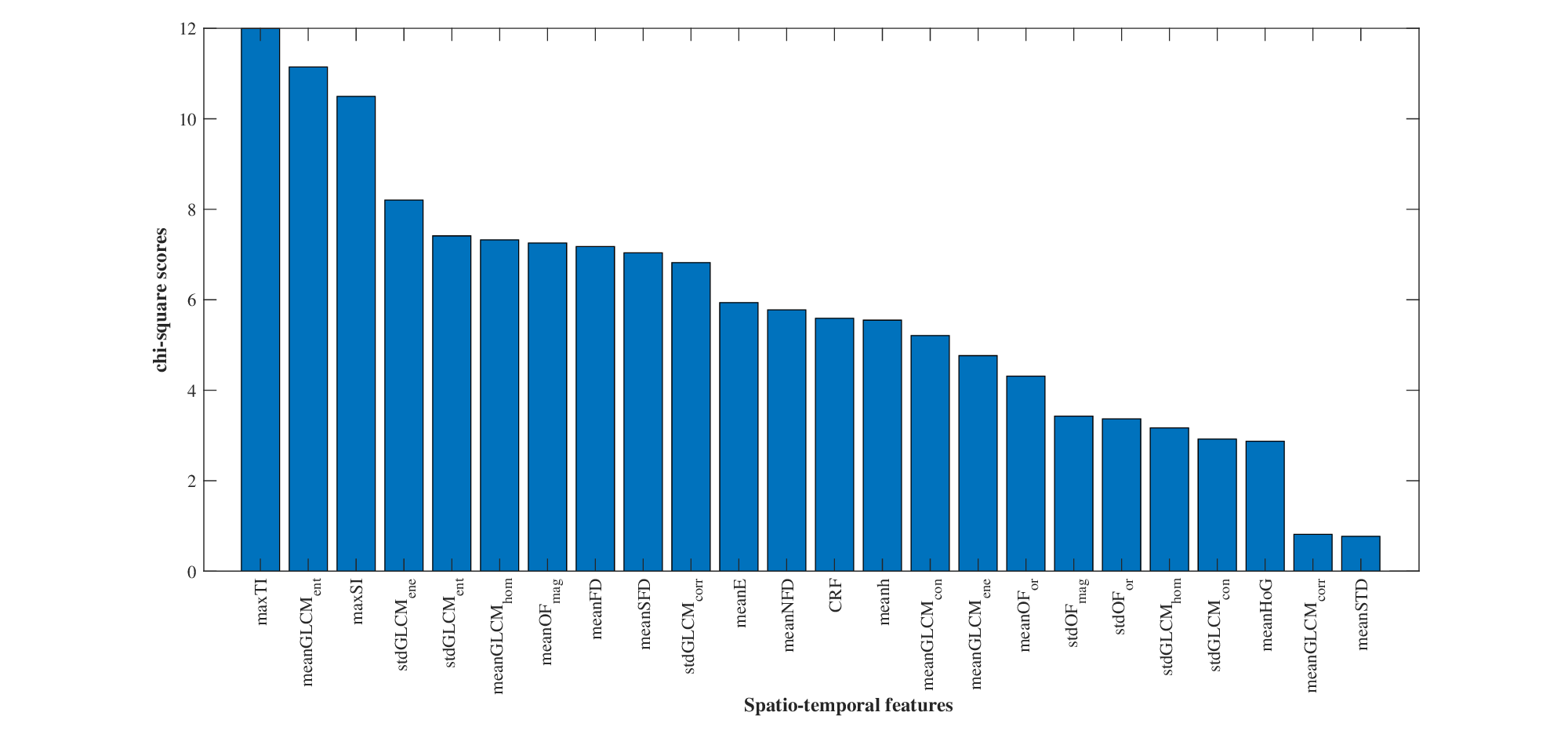}
\caption{Feature ranking of spatio-temporal features using chi-square scores. Only the spatio-temporal features with non-zero chi-square scores are shown here.}
\label{fig:featureSelection}
\end{figure*}

Figure \ref{fig:featureSelection} illustrates all the features from Table \ref{tab:spatiotemporalfeatures} with chi-square scores sorted in descending order. \textcolor{black}{Eventually, a subset of the most significant features was selected based on chi-square scores, as highlighted in Table \ref{tab:spatiotemporalfeatures}. The number of selected features, 15, was determined by identifying the smallest subset that achieves accuracy comparable to that of the entire feature space.}

\subsection{Training and Testing}
\label{subsec:TT}
We use the subset of features selected using feature ranking from Table \ref{tab:spatiotemporalfeatures} and labels as the optimal downsampling factors from Table \ref{tab:optFR} to train the supervised ensemble-based machine learning approach. We built an ensemble classifier using a decision tree as the base estimator. To combine the predictions of base classifiers, we use the bagging method \cite{baggingboosting}, which builds several instances of a base estimator on random subsets of the original training set and then combines their predictions to produce a final prediction. This method generates training data samples by randomly sampling from the training data. A base model is created on each of these samples for every iteration. These models run in parallel and are independent of each other. The final predictions are determined by combining the predictions from all the models. These models collectively form a higher-graded model to produce more accuracy. We use a 12-fold cross-validation method with random fold selection. With this technique, in each iteration, 80\% of the bit streams are used for training and 20\% for testing. After the twelfth iteration, the classification accuracy is averaged over all 12 iterations.

\subsection{Results and Discussion}
\label{subsec:results}
\textcolor{black}{The confusion matrix for the proposed frame rate selection method, EAFRS, is shown in Table \ref{tab:conf}. EAFRS achieves an overall classification accuracy of 92\%. Notably, the classifier correctly identifies bit streams with 120 fps and 60 fps as energy-aware frame rates with 100\% accuracy, without any misclassifications. For bit streams with ground truth frame rates of 30 fps, 24 fps, and 15 fps, occasional misclassifications occur, consistently toward higher frame rates. This behavior shows that no quality loss results from misclassifications.}

\begin{table}[]
\centering
\caption{Confusion Matrix of the proposed EAFRS method.}
\begin{tabular}{|c||ccccc|}
\hline
\multicolumn{1}{|l||}{} & \multicolumn{5}{c|}{\textbf{Predicted Frame Rate}} \\ \hline \hline
\textbf{Ground Truth} & \multicolumn{1}{c|}{120} & \multicolumn{1}{c|}{60} & \multicolumn{1}{c|}{30} & \multicolumn{1}{c|}{24} & 15 \\ \hline \hline 
120 & \multicolumn{1}{c|}{76} & \multicolumn{1}{c|}{0} & \multicolumn{1}{c|}{0} & \multicolumn{1}{c|}{0} & 0 \\ \hline
60 & \multicolumn{1}{c|}{0} & \multicolumn{1}{c|}{2} & \multicolumn{1}{c|}{0} & \multicolumn{1}{c|}{0} & 0 \\ \hline
30 & \multicolumn{1}{c|}{4} & \multicolumn{1}{c|}{1} & \multicolumn{1}{c|}{5} & \multicolumn{1}{c|}{0} & 0 \\ \hline
24 & \multicolumn{1}{c|}{1} & \multicolumn{1}{c|}{0} & \multicolumn{1}{c|}{0} & \multicolumn{1}{c|}{4} & 0 \\ \hline
15 & \multicolumn{1}{c|}{1} & \multicolumn{1}{c|}{0} & \multicolumn{1}{c|}{1} & \multicolumn{1}{c|}{1} & 16 \\ \hline
\end{tabular}
\label{tab:conf}
\end{table}

\begin{table}[]
\centering
\caption{Predicted frame rates obtained from the proposed EAFRS method (Section VI) for the subset of CRF values $c =\{18, 23, 28, 33\}$, with its associated BDR, BDEE, and BDDE values, which represents the average bitrate, encoding, and decoding energy savings for all the sequences and subset of sequences which is downsampled.}
\begin{tabular}{|l|lrrr|}
\hline
 & \multicolumn{4}{c|}{\textbf{EAFRS Method}} \\ \hline
\multirow{2}{*}{\textbf{File Name}} & \multicolumn{1}{l|}{\multirow{2}{*}{\begin{tabular}[c]{@{}l@{}}\textbf{Predicted energy-}\\ \textbf{aware frame rates}\end{tabular}}} & \multicolumn{3}{c|}{\textbf{BD (\%)}} \\ \cline{3-5}  & \multicolumn{1}{l|}{} & \multicolumn{1}{l|}{\textbf{BDR}} & \multicolumn{1}{l|}{\textbf{BDEE}} & \textbf{BDDE} \\ \hline
bobblehead & \multicolumn{1}{l|}{\{120,120,120,120\}} & \multicolumn{1}{r|}{0} & \multicolumn{1}{r|}{0} & 0 \\ \hline
books & \multicolumn{1}{l|}{\{120,120,120,120\}} & \multicolumn{1}{r|}{0} & \multicolumn{1}{r|}{0} & 0 \\ \hline
bouncyball & \multicolumn{1}{l|}{\{120,120,120,120\}} & \multicolumn{1}{r|}{0} & \multicolumn{1}{r|}{0} & 0 \\ \hline
catch & \multicolumn{1}{l|}{\{120,30,15,15\}} & \multicolumn{1}{r|}{-16.03} & \multicolumn{1}{r|}{-69.45} & -64.60 \\ \hline
catch\_track & \multicolumn{1}{l|}{\{120,120,120,120\}} & \multicolumn{1}{r|}{0} & \multicolumn{1}{r|}{0} & 0 \\ \hline
cyclist & \multicolumn{1}{l|}{\{120,120,120,120\}} & \multicolumn{1}{r|}{0} & \multicolumn{1}{r|}{0} & 0 \\ \hline
flowers & \multicolumn{1}{l|}{\{120,30,15,15\}} & \multicolumn{1}{r|}{16.43} & \multicolumn{1}{r|}{-52.73} & -52.61 \\ \hline
golf\_side & \multicolumn{1}{l|}{\{120,15,15,15\}} & \multicolumn{1}{r|}{-14.75} & \multicolumn{1}{r|}{-77.37} & -73.10 \\ \hline
guitar\_focus & \multicolumn{1}{l|}{\{120,120,30,24\}} & \multicolumn{1}{r|}{13.76} & \multicolumn{1}{r|}{-23.7} & -27.12 \\ \hline
hamster & \multicolumn{1}{l|}{\{120,120,120,120\}} & \multicolumn{1}{r|}{0} & \multicolumn{1}{r|}{0} & 0 \\ \hline
joggers & \multicolumn{1}{l|}{\{120,120,120,120\}} & \multicolumn{1}{r|}{0} & \multicolumn{1}{r|}{0} & 0 \\ \hline
lamppost & \multicolumn{1}{l|}{\{120,60,15,15\}} & \multicolumn{1}{r|}{20.04} & \multicolumn{1}{r|}{-32.08} & -32.58 \\ \hline
leaves\_wall & \multicolumn{1}{l|}{\{120,120,120,24\}} & \multicolumn{1}{r|}{-2.60} & \multicolumn{1}{r|}{-5.27} & -5.72 \\ \hline
library & \multicolumn{1}{l|}{\{120,120,120,30\}} & \multicolumn{1}{r|}{-0.72} & \multicolumn{1}{r|}{-4.58} & -4.52 \\ \hline
martial\_arts & \multicolumn{1}{l|}{\{120,120,120,120\}} & \multicolumn{1}{r|}{0} & \multicolumn{1}{r|}{0} & 0 \\ \hline
plasma & \multicolumn{1}{l|}{\{120,120,120,120\}} & \multicolumn{1}{r|}{0} & \multicolumn{1}{r|}{0} & 0 \\ \hline
pond & \multicolumn{1}{l|}{\{60,15,15,15\}} & \multicolumn{1}{r|}{-55.85} & \multicolumn{1}{r|}{-82.54} & -79.96 \\ \hline
pour & \multicolumn{1}{l|}{\{120,120,120,30\}} & \multicolumn{1}{r|}{0.21} & \multicolumn{1}{r|}{-3.57} & -4.86 \\ \hline
sparkler & \multicolumn{1}{l|}{\{120,120,120,120\}} & \multicolumn{1}{r|}{0} & \multicolumn{1}{r|}{0} & 0 \\ \hline
typing & \multicolumn{1}{l|}{\{120,120,24,15\}} & \multicolumn{1}{r|}{1.09} & \multicolumn{1}{r|}{-30.26} & -33.09 \\ \hline
water\_ripples & \multicolumn{1}{l|}{\{120,120,24,24\}} & \multicolumn{1}{r|}{16.27} & \multicolumn{1}{r|}{-17.21} & -19.11 \\ \hline
water\_splashing & \multicolumn{1}{l|}{\{120,120,120,120\}} & \multicolumn{1}{r|}{0} & \multicolumn{1}{r|}{0} & 0 \\ \hline
Beauty & \multicolumn{1}{l|}{\{120,120,120,120\}} & \multicolumn{1}{r|}{0} & \multicolumn{1}{r|}{0} & 0 \\ \hline
Bosphorus & \multicolumn{1}{l|}{\{120,120,120,30\}} & \multicolumn{1}{r|}{-1.69} & \multicolumn{1}{r|}{-9.79} & -9.70 \\ \hline
HoneyBee & \multicolumn{1}{l|}{\{60,15,15,15\}} & \multicolumn{1}{r|}{-71.00} & \multicolumn{1}{r|}{-84.44} & -82.06 \\ \hline
Jockey & \multicolumn{1}{l|}{\{120,120,120,120\}} & \multicolumn{1}{r|}{0} & \multicolumn{1}{r|}{0} & 0 \\ \hline
ReadySteadyGo & \multicolumn{1}{l|}{\{120,120,120,120\}} & \multicolumn{1}{r|}{0} & \multicolumn{1}{r|}{0} & 0 \\ \hline
YachtRide & \multicolumn{1}{l|}{\{120,120,120,120\}} & \multicolumn{1}{r|}{0} & \multicolumn{1}{r|}{0} & 0 \\ \hline
\textbf{All sequences} & \multicolumn{1}{l|}{\textbf{Average BD}} & \multicolumn{1}{r|}{\textbf{-3.38}} & \multicolumn{1}{r|}{\textbf{-17.46}} & \textbf{-17.60} \\ \hline
\textbf{Downsampled} & \multicolumn{1}{l|}{\textbf{Average BD}} & \multicolumn{1}{r|}{\textbf{-5.58}} & \multicolumn{1}{r|}{\textbf{-29.00}} & \textbf{-28.76} \\ 
\textbf{sequences} & \multicolumn{1}{l|}{} & \multicolumn{1}{r|}{} & \multicolumn{1}{r|}{} &  \\ \hline
\end{tabular}
\label{tab:predFR}
\end{table}

\textcolor{black}{The average BDR, BDEE, and BDDE values obtained by the frame rate recommendation from the EAFRS method (trained model), are tabulated in Table \ref{tab:predFR}. The average values for the entire set of sequences are -3.38\%, -17.46\%, and -17.60 \%, respectively. When considering only the sequences for which downsampling leads to savings, the BDR, BDEE, and BDDE values are -5.58\%, -29\%, and -28.76\%, respectively. We can see that the values obtained by the EAFRS method are close to those at optimal frame rates (-4.41\%, -32.03\%, and -32.17\%, respectively), despite misclassifications.}

\subsection{Benchmarking}
\label{subsec:becnhmarking}
\textcolor{black}{We benchmark the proposed EAFRS method against the approach presented in \cite{Bulletal18}. Both methods employ machine learning-based classifiers; however, the training conditions and ground truth definitions differ. In \cite{Bulletal18}, the ground truth frame rates are determined based purely on optimizing video quality using Differential Mean Opinion Scores (DMOS). In contrast, in our method, the ground truth frame rates are selected based on Pareto-optimal trade-offs between energy consumption and visual quality, rather than solely on visual quality.}

\textcolor{black}{We reproduce the method from \cite{Bulletal18} and evaluate the frame rate predictions of both methods on the BVI-HFR dataset (22 sequences). The corresponding results are presented in Table \ref{tab:compare}. The ground truths and the training feature subsets differ between the two methods. Therefore, the observed differences in energy and bitrate savings reflect the different optimization targets (quality-optimal versus energy-aware). Moreover, the method in \cite{Bulletal18} does not account for compression, whereas the proposed method explicitly considers the quantization setting (CRF) to identify optimal downsampling factors.}

\textcolor{black}{Table \ref{tab:compare} shows that, compared to the method from \cite{Bulletal18}, the proposed EAFRS method achieves greater energy savings (e.g., a BDEE of -18.13\% compared to -10.69\%) and additionally provides bitrate savings, whereas the method from \cite{Bulletal18} results in an increased bitrate demand. In summary, the results demonstrate that optimizing frame rates based on energy and quality trade-offs, rather than solely on quality, can lead to significant energy and bitrate savings while maintaining quality.}

\begin{table}[]
\centering
\caption{BDR, BDEE, and BDDE values of expected (ground truth) and predicted frame rates of the BVI-HFR dataset for our proposed method and method from \cite{Bulletal18}.}
\begin{tabular}{|l|l|lll|}
\hline
\multirow{2}{*}{\textbf{Method}} & \multirow{2}{*}{\textbf{Savings}} & \multicolumn{3}{c|}{\textbf{BD Metric (\%)}} \\ \cline{3-5} 
 &  & \multicolumn{1}{l|}{\textbf{BDR}} & \multicolumn{1}{l|}{\textbf{BDEE}} & \textbf{BDDE} \\ \hline
\multirow{2}{*}{\textbf{Proposed EAFRS method}} & Ground Truth & \multicolumn{1}{l|}{-0.75} & \multicolumn{1}{l|}{-19.19} & -19.26 \\ \cline{2-5} 
 & Prediction & \multicolumn{1}{l|}{-1.01} & \multicolumn{1}{l|}{-18.13} & -18.05 \\ \hline
\multirow{2}{*}{\cite{Bulletal18}} & Ground Truth & \multicolumn{1}{l|}{8.67} & \multicolumn{1}{l|}{-12.45} & -12.12 \\ \cline{2-5} 
 & Prediction & \multicolumn{1}{l|}{8.16} & \multicolumn{1}{l|}{-10.69} & -10.98 \\ \hline
\end{tabular}
\label{tab:compare}
\end{table}

\subsection{Complexity Analysis}
\label{subsec:complexity}
\textcolor{black}{We evaluate the computational complexity of the proposed EAFRS method in terms of the energy overhead it introduces. To do so, we compute the relative energy consumption difference, denoted as  $\Delta E_{\text{select}}$, across all considered CRF values by comparing the following two encoding scenarios:
\begin{itemize}
    \item[(a)] Encoding at an energy-aware frame rate selected by EAFRS for each CRF, including the overhead of spatio-temporal feature extraction and classification.
    \item[(b)] Encoding each sequence for all the CRFs at the native frame rate (120 fps).
\end{itemize}}

\textcolor{black}{Table \ref{tab:complexity}, column 1, reports the $\Delta E_{\text{select}}$ for all the sequences. The negative values indicate energy savings achieved by selecting a lower frame rate, for the sequences with low-motion or temporally redundant content or both (e.g., catch, pond, HoneyBee), where the energy consumed for feature extraction, classification, and encoding is lower than that for constant 120 fps (native frame rate) encoding. The positive values, on the other hand, correspond to high-motion or high-detail sequences, or both (e.g., bobblehead, cyclist, Jockey), where feature extraction and classification incur an additional energy cost without reducing frame rate. Such cases occur when EAFRS retains the original frame rate for all the CRF values to preserve visual quality. However, the observed increases are minimal and do not significantly affect the overall efficiency of the method. On average, the proposed method yields a relative energy saving of 17.30\% across all sequences, with an average energy saving of 38.43\% for sequences where the proposed EAFRS reduces the frame rate and an average additional energy overhead of 1.01\% for sequences where the original frame rate is maintained.}

\textcolor{black}{In summary, these results confirm that EAFRS substantially reduces encoding energy consumption while introducing only negligible computational overhead, primarily for sequences where temporal downsampling does not lead to energy savings.}

\begin{table}[]
\centering
\caption{Energy differences between encoding with frame rates selected by the proposed method, including the feature extraction and classification, and encoding at the native frame rate for all evaluated sequences.}
\label{tab:complexity}
\begin{tabular}{|l|r|}
\hline
\textbf{Sequence} & $\mathbf{\Delta E_{\text{select}}}$ \textbf{(\%)} \\ \hline
bobblehead       & 0.61   \\ \hline
books            & 1.18   \\ \hline
bouncyball       & 0.78   \\ \hline
catch            & -54.85 \\ \hline
catch\_track     & 0.94   \\ \hline
cyclist          & 1.15   \\ \hline
flowers          & -51.65 \\ \hline
golf\_side       & -58.22 \\ \hline
guitar\_focus    & -30.97 \\ \hline
hamster          & 0.97   \\ \hline
joggers          & 1.21   \\ \hline
lamppost         & -40.35 \\ \hline
leaves\_wall     & -16.30 \\ \hline
library          & -14.74 \\ \hline
martial\_arts    & 1.35   \\ \hline
plasma           & 0.84   \\ \hline
pond             & -72.20 \\ \hline
pour             & -11.92 \\ \hline
sparkler         & 0.74   \\ \hline
typing           & -33.98 \\ \hline
water\_ripples   & -29.22 \\ \hline
water\_splashing & 0.43   \\ \hline
Beauty           & 1.27   \\ \hline
Bosphorus        & -12.57 \\ \hline
HoneyBee         & -72.57 \\ \hline
Jockey           & 1.22   \\ \hline
ReadySteadyGo    & 1.32   \\ \hline
YachtRide        & 1.15   \\ \hline\hline
\textbf{Average} & \textbf{-17.30}  \\ \hline
\end{tabular}
\end{table}

\section{Conclusion}
\label{sec:conclusion}
\textcolor{black}{This work demonstrates that temporal downsampling, when applied adaptively based on content characteristics and quantization levels, is more effective for compression and energy efficiency than using a fixed frame rate for all sequences and CRF values. Our findings are validated through a subjective evaluation of energy-aware frame rate and CRF pairs, demonstrating a strong correlation between objective quality metrics and subjective opinion scores. }

\textcolor{black}{In addition, we introduced an Energy-Aware Frame Rate Selection (EAFRS) method, which extracts spatio-temporal features from video sequences and employs a feature-based supervised machine learning approach to predict energy-aware frame rates for a given CRF. The proposed EAFRS method achieves a 92\% accuracy in predicting the optimal energy-aware frame rate and delivers significant energy savings, with an average reduction of 3.38\% in bitrate, 17.46\% in encoding energy, and 17.60\% in decoding energy, all while preserving video quality.}

\textcolor{black}{For future work, we aim to generalize the proposed method for video sequences with varied source frame rates, ensuring its applicability across a broader range of content. Additionally, we plan to expand the subjective evaluation by testing a larger dataset with more frame rates, further refining our approach. Another key direction is to extend EAFRS to support variable frame rates, adapting dynamically as video content changes over time.} In addition, we plan to investigate the spatiotemporal adaptation of video sequences for energy efficiency.

\bibliographystyle{IEEEtran}
\bibliography{literatureNeu}
\vspace{-0.01cm}
\begin{IEEEbiography}[{\includegraphics[width=1in,height=1.25in,clip,keepaspectratio]{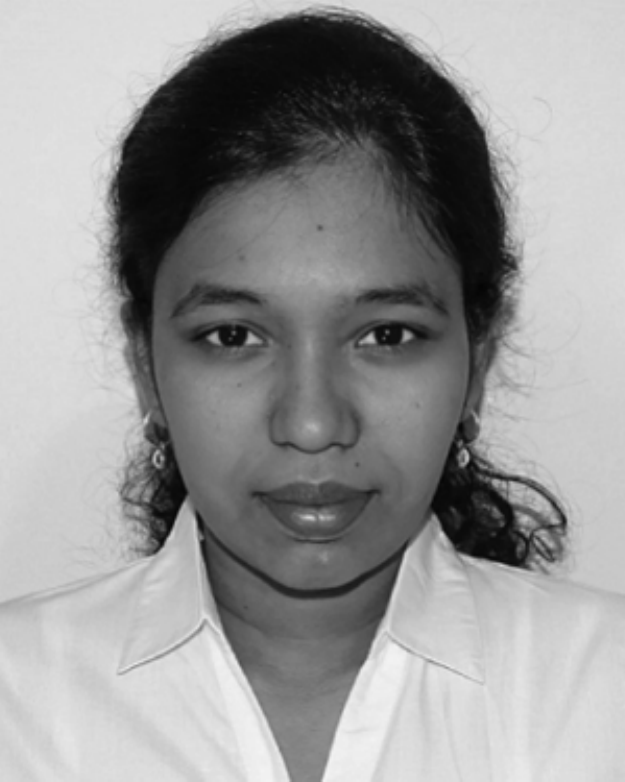}}] {Geetha Ramasubbu}(Graduate Student Member,
IEEE) received the B.E. degree in electronics and
communication engineering from Anna University,
Chennai, India, in 2015. She studied communications and multimedia engineering from Friedrich Alexander University Erlangen-Nürnberg (FAU),
Germany in 2017, and graduated with a M.Sc. degree in 2020. From 2015 to 2017, she worked as a Software Developer for Cognizant Technology Solutions, India. Since 2021, she has been a Research Scientist with the Chair of Multimedia Communications and Signal Processing, FAU. Her research interests include energy-efficient video communications and video coding.
\end{IEEEbiography}

\begin{IEEEbiography}[{\includegraphics[width=1in,height=1.25in,clip,keepaspectratio]{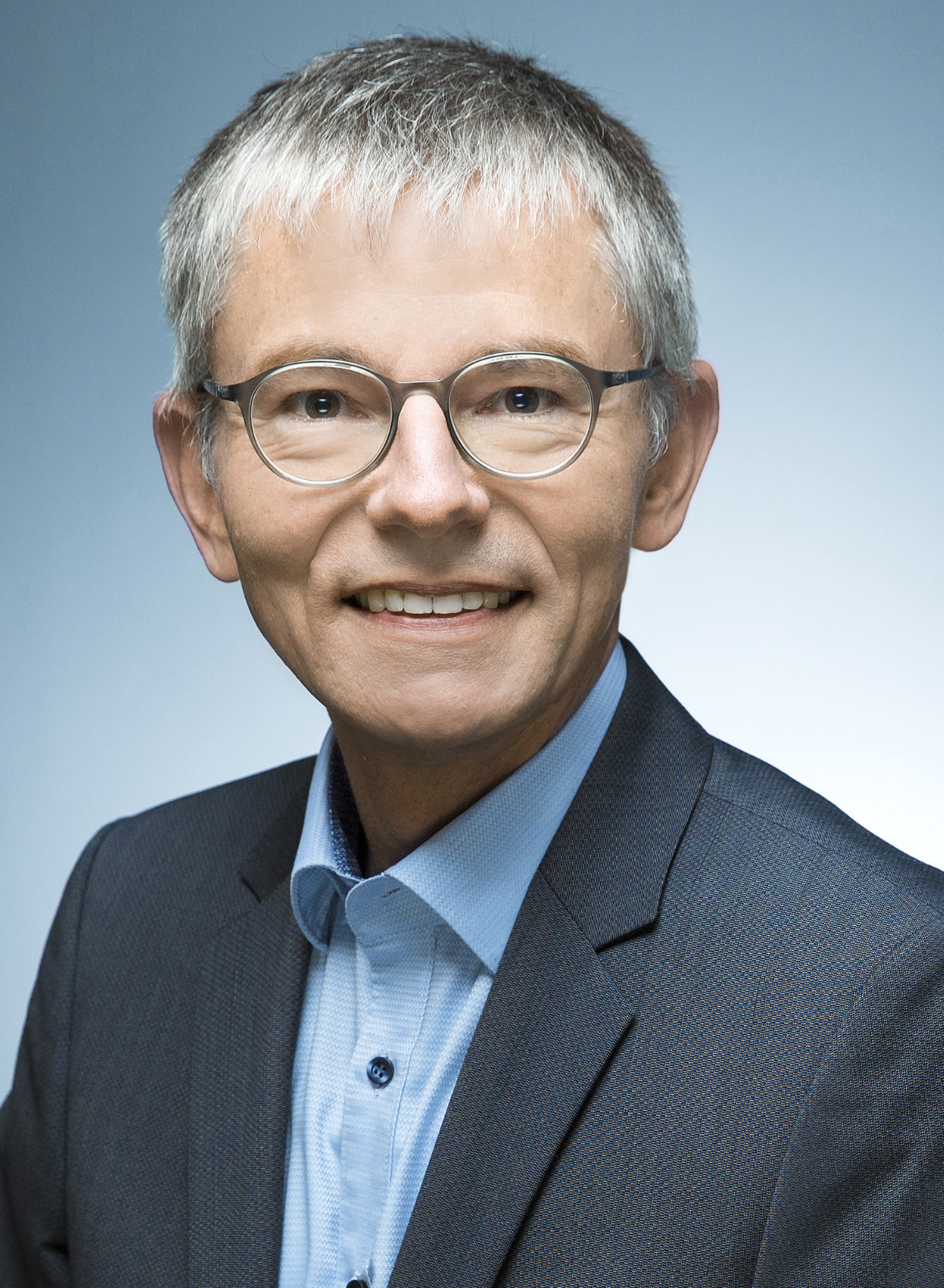}}]{Andr\'e Kaup} (Fellow, IEEE) received the Dipl.-Ing. and Dr.-Ing. degrees in electrical engineering from RWTH Aachen University, Aachen, Germany, in 1989 and 1995, respectively. He joined Siemens Corporate Technology, Munich, Germany, in 1995, and became the Head of the Mobile Applications and Services Group in 1999. Since 2001, he has been a Full Professor and the Head of the Chair of Multimedia Communications and Signal Processing at Friedrich-Alexander University Erlangen-Nürnberg (FAU), Germany. From 2005 to 2007 he was Vice Speaker of the DFG Collaborative Research Center 603. From 2015 to 2017, he served as the Head of the Department of Electrical Engineering and Vice Dean of the Faculty of Engineering at FAU. He has authored around 500 journal and conference papers and has over 120 patents granted or pending. His research interests include image and video signal processing and coding, and multimedia communication.
Dr. Kaup is vice-chair of the IEEE Image, Video, and Multidimensional Signal Processing Technical Committee and a member of the Scientific Advisory Board of the German VDE/ITG. He is an IEEE Fellow and a member of the Bavarian Academy of Sciences and Humanities, the German National Academy of Science and Engineering, and the European Academy of Sciences and Arts. He is a member of the Editorial Board of the IEEE Circuits and Systems Magazine. He was a Siemens Inventor of the Year 1998 and obtained the 1999 ITG Award. He received several IEEE best paper awards, including the Paul Dan Cristea Special Award in 2013, and his group won the Grand Video Compression Challenge from the Picture Coding Symposium 2013. The Faculty of Engineering with FAU and the State of Bavaria honored him with Teaching Awards, in 2015 and 2020, respectively. He served as an Associate Editor of the IEEE Transactions on Circuits and Systems for Video Technology. He was a Guest Editor of the IEEE Journal of Selected Topics in Signal Processing.\end{IEEEbiography}

\begin{IEEEbiography}[{\includegraphics[width=1in,height=1.25in,clip,keepaspectratio]{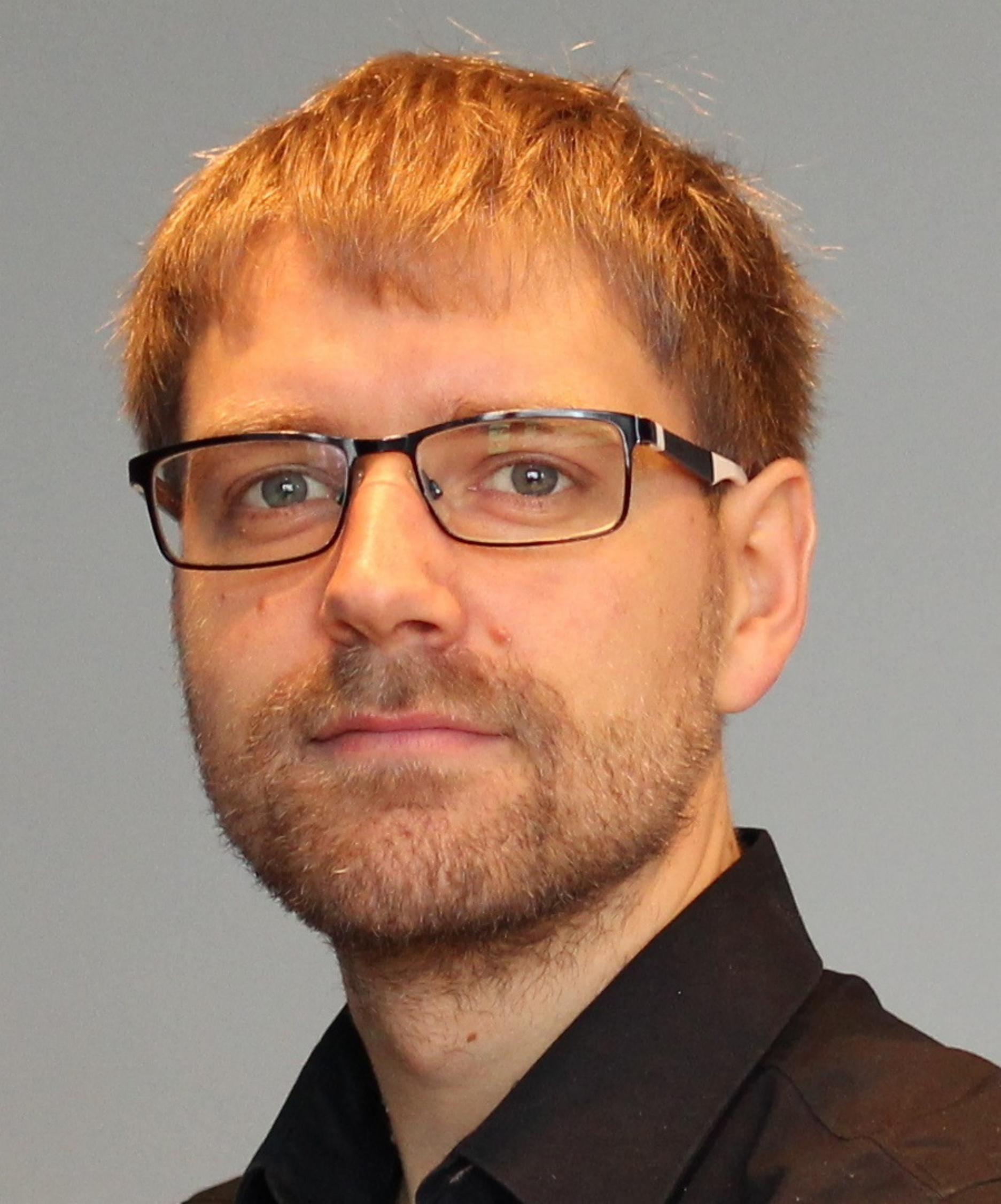}}]{Christian Herglotz} (Member, IEEE) received
the Dipl.-Ing. degree in electrical engineering
and information technology and the Dipl.-Wirt.-
Ing. degree in business administration and economics from Rheinisch-Westfälische Technische
Hochschule (RWTH) Aachen University, Germany,
in 2011 and 2012, respectively, and the Dr.-Ing.
degree from the Chair of Multimedia Communications and Signal Processing, Friedrich-Alexander Universität Erlangen-Nürnberg (FAU), Germany,
in 2017.
From 2012 to 2023, he was a Research Scientist with the Chair of
Multimedia Communications and Signal Processing, FAU. From 2018 to 2019,
he was a Postdoctoral Fellow with École de technologie supérieure in
collaboration with Summit Tech Multimedia, Montreal, Canada, on energy
efficient VR technologies. Since 2023, he has been a Substitute Professor of
computer engineering with Brandenburgische Technische Universität Cottbus Senftenberg, Germany. His current research interests include energy efficient
video communications, video coding, and efficient hardware and software
implementations for image and video processing. Since 2020, he has been
with the Visual Signal Processing and Communications Technical Committee
of the IEEE Circuits and Systems Society. From 2023-2024, he served as an Associate Editor for the IEEE Transactions on Circuits and Systems for Video Technology.
\end{IEEEbiography}
\EOD
\end{document}